\documentclass{iaus}
\usepackage{graphicx}

\title[Of?p stars: A class of slowly rotating magnetic massive stars] %% give here short title %%
{Of?p stars: A class of slowly rotating magnetic massive stars}

\author[G.A. Wade et al.]   %% give here short author list %%
{G.A. Wade$^1$, J.H. Grunhut$^{1}$, W.L.F. Marcolino$^2$, F. Martins$^3$, \\I. Howarth$^4$, Y. Naz\'e$^5$, N. Walborn$^6$ and the MiMeS Collaboration}

\affiliation{$^1$RMC, Canada; $^2$ONB, Brazil; $^3$GRAAL, France; $^4$UCL, UK; $^5$Li\`ege, Belgium; $^6$STScI, USA\\}

\pubyear{2010}
\volume{272}  %% insert here IAU Symposium No.
\pagerange{119--126}
\jname{Active OB stars}
\editors{A.C. Editor, B.D. Editor \& C.E. Editor, eds.}
\begin{document}

\maketitle

\begin{abstract}
Only 5 Of?p stars have been identified in the Galaxy. Of these, 3 have been studied in detail, and within the past 5 years magnetic fields have been detected in each of them. The observed magnetic and spectral characteristics are indicative of organised magnetic fields, likely of fossil origin, confining their supersonic stellar winds into dense, structured magnetospheres. The systematic detection of magnetic fields in these stars strongly suggests that the Of?p stars represent a general class of magnetic O-type stars.
\end{abstract}

\firstsection % if your document starts with a section,
\section{Introduction}
The enigmatic Of?p stars are identified by a number of peculiar and outstanding observational properties. The classification was first introduced by Walborn (1972) according to the presence of C~{\sc iii} $\lambda 4650$ emission with a strength comparable to the neighbouring N~{\sc iii} lines. Well-studied Of?p stars are now known to exhibit recurrent, and apparently periodic, spectral variations (in Balmer, He~{\sc i}, C~{\sc iii} and Si~{\sc iii} lines) with periods ranging from days to decades, strong C~{\sc iii} $\lambda 4650$ in emission, narrow P Cygni or emission components in the Balmer lines and He~{\sc i} lines, and UV wind lines weaker than those of typical Of supergiants (see Naz\'e et al. 2010 and references therein). 

Only 5 Galactic Of?p stars are known (Walborn et al. 2010): HD 108, HD 148937, HD 191612, NGC 1624-2 and CPD$-28^{\rm o} 2561$. Three of these stars - HD 108, HD 148937 and HD 191612 - have been studied in detail. In recent years, HD 191612 was  carefully examined for the presence of magnetic fields (Donati et al. 2006), and was clearly detected. Recent observations, obtained chiefly within the context of the Magnetism in Massive Stars (MiMeS) Project (Martins et al. 2010; Wade et al., in prep) have furthermore detected magnetic fields in HD 108 and HD 148937, thereby confirming the view of Of?p stars as a class of slowly rotating, magnetic massive stars.

%\begin{figure} , as shown in Fig.~\ref{lsd_examp},
%\centering
%\includegraphics[width=1.8in]{hd142184_03mar10pn.s_lsd_norm.eps}
%\caption{Example mean LSD Stokes~$V$ (top), diagnostic null (middle), and unpolarized Stokes~$I$ line profiles of HD~142184.}
%\label{lsd_examp}
%\end{figure}

\section{HD 191612}

HD 191612 was the first Of?p star in which a magnetic field was detected (Donati et al. 2006). Subsequent MiMeS observations with ESPaDOnS@CFHT (Wade et al., in prep) confirm the existence of the field, and demonstrate the sinusoidal variability of the longitudinal field with the H$\alpha$ and photometric period of 537.6 d. As shown in Fig. 1, the longitudinal field, H$\alpha$ and photometric extrema occur simultaneously when folded according to the 537.6 d period. This implies a clear relationship between the magnetic field and the circumstellar envelope. We interpret these observations in the context of the oblique rotator model, in which the stellar wind couples to the kilogauss dipolar magnetic field, generating a dense, structured magnetosphere, resulting in all observables varying according to the stellar rotation period.
\begin{figure}
\centering
\includegraphics[width=3in,angle=-90]{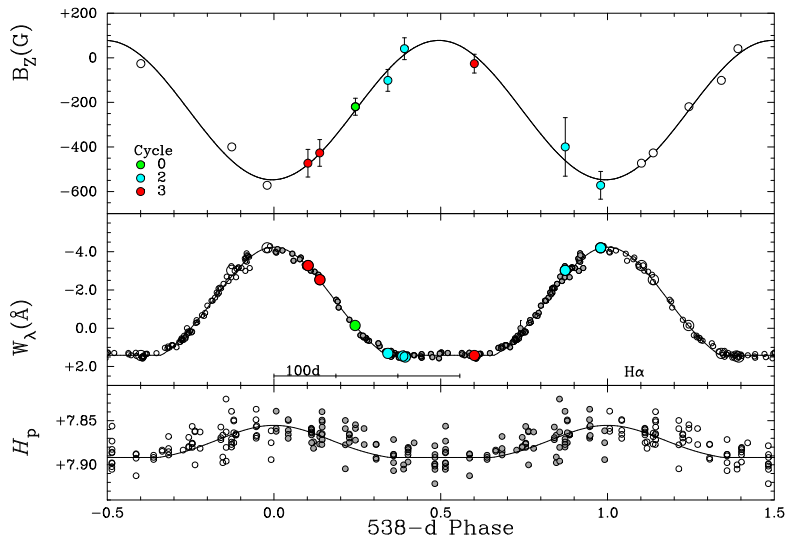}
\caption{Longitudinal field (top), H$\alpha$ EW (middle) and Hipparcos mag (bottom) of the Of?p star HD 191612, all phased according to the 537.6 d period. From Wade et al., in preparation.}
\label{mag_geo}
\end{figure}

\section{HD 108}

HD 108 was the second Of?p star in which a magnetic field was detected (Martins et al. 2010). Based on long-term photometric and spectroscopic monitoring, HD 108 is suspected to vary on a timescale of ~50-60 y (Naz\'e et al. 2001). The magnetic observations acquired by Martins et al. from 2007-2009 show at most a marginal increase of the longitudinal field during more than 2 years of observation. This supports the proposal that the variation timescale is in fact the stellar rotational period, and that HD 108 is a magnetic oblique rotator that has undergone extreme magnetic braking.

\section{HD 148937}

HD 148937 was recently observed intensely by the MiMeS Collaboration, resulting in the detection of circular polarisation within line profiles indicative of the presence of an organised magnetic field of kilogauss strength (Wade et al., in prep). Although the field is consistently detected in the observations, no variability is observed, in particular according to the 7.03 d spectral period. This result supports the proposal by Naz\'e et al. (2010) that HD 148937 is observed with our line-of-sight near the stellar rotational pole.

\end{document}